\documentstyle[proceedings,numreferences]{crckapb}
\begin{opening}
\title{Resonant Tunneling and Charging Effects, a Path Integral Approach}
\author{J\"urgen K\"onig$^1$}
\author{Herbert Schoeller$^{1,2}$}
\author{Gerd Sch\"on$^1$}
\author{Rosario Fazio$^{1,3}$}
\institute{
$^1$ Institut f\"ur Theoretische Festk\"orperphysik, \\
Universit\"at Karlsruhe, 76128 Karlsruhe, Germany\\
$^2$ Department of Physics, Simon Fraser University, Burnaby, B.C.,
V5A 1S6, Canada\\
$^3$ Istituto di Fisica, Facolta di Ingegneria, 95129 Catania, Italy}
\runningauthor{J. K\"onig, H. Schoeller, G. Sch\"on, R. Fazio}
\end{opening}

\begin{document}

\begin{abstract}
Electron tunneling through small metallic islands with low
capacitance is studied.  The large charging energy in these systems
is responsible for nonperturbative Coulomb blockade effects.
We further consider the effect of electron interactions in the electrodes.
In junctions with high resistance compared to the quantum resistance
transport can be described by sequential tunneling.
If the resistance is lower, quantum fluctuations,
higher order coherent processes,
and eventually resonant tunneling become important.
We present a path integral real-time approach, which
allows a systematic diagrammatic classification of these processes.
An important process is ``inelastic resonant tunneling'',
where different electrons tunnel coherently between
the electrodes and the island.
Physical quantities like the current and the average charge on the
island can be deduced.
We find a strong renormalization of
the system parameters and, in addition, a finite lifetime broadening.
It results in a pronounced broadening and smearing
of the Coulomb oscillations of the conductance. These effects
are important in an experimentally accessible range of temperatures.
The electron interaction in the electrodes is modeled by a Luttinger liquid.
It leads to non-analytic kernels in the effective action.
The diagrammatic expansions can be performed also in this case, resulting in
power-law current-voltage characteristics.
\end{abstract}

\section{Introduction}

Electron transport through mesoscopic metallic islands
coupled to electrodes has been the subject of extensive
research [1-3].
The small size and capacitance of these systems imply a strong Coulomb
interaction energy $E_{ch}$, which gives rise to a variety of single-electron
phenomena. At low temperatures
tunneling can be suppressed  (Coulomb blockade).
The charging energy can be tuned by gate voltages.
When the energy difference between adjacent charge states
$E_{ch}(n\pm1)-E_{ch}(n)$ is lower than the temperature $kT$ or
the bias voltage $eV$, a current flows through the system.
As a consequence, the conductance shows
a series of peaks as a function of the gate voltage $V_G$ (linear
response) and further structure at larger bias voltage $V$
(nonlinear response).
The classical description is sufficient as long as the resistance
$R_T$ of a single barrier is much higher than the quantum resistance
$R_K= h/e^2 $, i.e. for $\alpha_0 \ll1$, where
$\alpha_0\equiv R_K/(4\pi^2 R_T)  \; .$
In this regime, transport occurs in
sequences of uncorrelated tunneling processes. The rates can be obtained
in lowest order perturbation theory in the tunneling amplitudes.
They enter a master equation
[4-8], from which the probabilities for different charge states $n$
and the currents can be calculated.

For very low temperatures or when the dimensionless conductance
$\alpha_0$ is not small the classical description breaks down. Quantum
fluctuations and higher order coherent tunneling processes become important
[9-15].
This includes cotunneling, where two electrons
tunnel coherently in different junctions, thus avoiding the Coulomb blockade.
Furthermore, resonant tunneling,
where electrons tunnel coherently back and forth between the island and
the electrodes,  play a role. A description is called for, which allows a
systematic classification of all these processes.
In comparison to the well known phenomenon of resonant tunneling of
single electrons one encounters here two complications.
One lies in the fact that
the metallic system contains many electrons. With overwhelming
probability different electron states are involved in the different transitions
of the coherent process. The second arises since the
Coulomb interaction is strong and, hence, cannot be accounted for in
perturbation theory.

In the present article we develop a systematic diagrammatic
technique to identify the processes of sequential
tunneling, inelastic cotunneling and resonant tunneling.
We study the time evolution of the density
matrix. In an earlier paper \cite{SS} we have formulated the problem,
after a separation of charge and fermionic degrees of freedom,
in a many-body expansion technique. Here we reformulate it in a real-time
path-integral representation. The latter is well known from the
studies of dissipation in quantum mechanics. Caldeira and Leggett
\cite{CLeggett}, following methods pioneered by Feynman and Vernon
\cite{Fey-Ver}, studied the problem of Ohmic dissipation.
Dissipation associated with tunneling of electrons
has been investigated in Refs. \cite{Eck-Scho-Amb,Schoen-Zai}.
An essential step in the present work is a transformation of the
Feynman-Vernon functional for electron tunneling
from a phase (coordinate) representation to a charge
(momentum) representation \cite{Schoen-Zai}.
The path-integral method is well-suited to account for the strong correlations
due to the Coulomb interaction in a perturbative or nonperturbative analysis
in the tunneling. In contrast, usual Green's
function techniques cannot be used since the Coulomb interaction has to
be retained in the unperturbed Hamiltonian.
The same problem arises in the context of local, strongly
correlated Fermi systems like the Kondo and Anderson model
\cite{Bic,Hew}. For these systems diagrammatic
techniques have been derived by Barnes \cite{Bar}
from a slave-boson description. Rammer \cite{Ram} developed a
graphical density-matrix description for the dynamics
of a particle coupled to a heat bath. The problem can also
be formulated in terms of Liouville operators \cite{Los-Schoeller}.

As examples we will study the electron box
and single electron transistors. The box
consists of a metallic island coupled via a tunnel junction with
capacitance $C_J$ to an electrode. It is further coupled
capacitively ($C_G$) to a voltage source $V_G$. The charging energy is
$E_{ch}(n) = (ne-Q_G)^2/(2C)$, where $C = C_J + C_G$
denotes the total capacitance of the island. It
depends on the number of excess electrons $n$ on the island and on
the continuously varying external charge $Q_G = C_G V_G$.
The single electron transistor consists of a metallic island which is
coupled by two tunnel junctions to two electrodes (see Fig.
\ref{fig1}). Here
a transport voltage $V = V_L - V_R$ drives a current.
The island is further coupled capacitively to a gate voltage $V_G$.
The charging energy of this system  depends again on the number of electrons
$n$ on the island, $E_{ch}(n) = (ne-Q_G)^2/(2C)$.
Here $C = C_L + C_R + C_G$ is the
sum of the two junctions and the gate capacitances
and $Q_G = C_G V_G + C_L V_L + C_R V_R$.
In tunneling processes we further have to account for the work done by the
voltage source (see below). In both examples the total island capacitance
defines the scale for the charging energy $E_C \equiv e^2/2C$.

\begin{figure}
\vspace{4.5cm}
\caption{
Equivalent circuit for the SET-transistor.
}
\label{fig1}
\end{figure}

The tunneling is described by tunneling  Hamiltonians. We consider
``wide'' metallic junctions,
which implies that there are many transverse channels.
As a result ``inelastic'' higher order tunneling processes,
involving different electron
states for each step, dominate over those higher order processes which
involve the same state repeatedly.
Accordingly, in the effective action description
presented below only simple loop diagrams are retained.

At low temperature  the electron number $n(Q_G)$ in the electron box
takes the value which minimizes the charging energy.
It increases in steps of unity when $Q_G$ is increased.
Tunneling processes in the transistor are only possible in lowest order
perturbation theory if the electrochemical potential of one
electrode is high enough to allow one electron to enter the island,
say $eV_L > E_{ch}(n+1) - E_{ch}(n) $, while the electrochemical
potential of the other electrode allows the next
tunneling process to that electrode, i.e.
$E_{ch}(n+1) - E_{ch}(n) > eV_R$.
Within the window set by these two conditions the current  is
$4R_t I = V - 4 [Q_G-(n+1/2)e]^2/(C^2V)$.
This implies that the conductance $G(Q_G) = \partial{I}/\partial{V}$
shows, as a function of $Q_G$, an $e$-periodic series of structures
of width $CV$ with, at $T=0$, vertical steps at its limits.
At finite temperature the average value $\langle n(Q_G) \rangle $ and
the steps in the linear and nonlinear conductance are washed out.

Quantum fluctuations further wash out these steps,
even at zero temperature.
We will describe these processes diagrammatically.
In the most interesting case we can re-sum the diagrams
and obtain closed expressions for the stationary density matrix and
the spectral density describing the charge excitations of the system.
Our main findings are:
(i) After a resummation of the leading logarithmic terms
in $\alpha_0 \ln{(E_C/ |\omega|)}$ we find a
renormalization of the energy and dimensionless conductance.
(ii) The coherent processes where electrons tunnel an
arbitrary number of times between the leads and the island (resonant
tunneling) give rise to a broadening of the charge state levels.
Both effects need to be retained in a conserving theory
which obeys sum rules and current conservation.
Both effects are observable in a real experiment at accessible temperatures.

In order to study the effect of electron interactions we consider
in the last section a model where the electrodes
are assumed to be Luttinger liquids \cite{Haldane}.
Due to the interactions the single particle
density of states has a power-law asymptotic at low energies, and Fermi
liquid theory does not apply anymore.
For a quantum wire with an arbitrarily small barrier this leads to a
suppression of transport at low energies~\cite{Kane,Matveev2}.
We find that the effective action in the
Luttinger liquid - normal metal junction
has the same form as in the metallic case. However, the kernels are modified,
taking a power law form with exponents depending on the
interaction strength in the Luttinger liquid.
The diagrammatic expansions can be performed and re-summed also in this case.
The resulting $I$-$V$-characteristics show a power-law dependence.

\section{Real time evolution of the density matrix}

The description of the single electron transistor
is based on the Hamiltonian
\begin{equation}
	H=H_L+H_R+H_I+H_{ch}+H_t \; .
\label{ham}
\end{equation}
Here
$H_L =\sum_{k,\sigma} (\epsilon_k + eV_L) c^{\dagger}_{k\sigma} c_{k\sigma}$
describes noninteracting electrons in the left lead, with similar expressions
for the island (with states denoted by $q$)  and the right lead.
The electrodes are treated as reservoirs, i.e. they
remain in thermal equilibrium. The Coulomb
interaction $H_{ch}$ is assumed to depend only on the total charge on
the island, as expressed by the charging energy  $E_{ch}(n)$ introduced above.
Charge transfer is described by the standard tunneling
Hamiltonian, e.g. the tunneling in the left junction by
\begin{equation}
	H_{t,L}=\sum_{k,q,\sigma} T_{kq}
	c^{\dagger}_{k\sigma} c_{q\sigma} + h.c. \; .
\label{5a}
\end{equation}
The tunnel matrix elements $T_{kq}$ are considered
independent of the states $k$ and $q$.
They can be related to the tunnel restistances $R_{L/R}$ of the junctions,
$1/R_{L/R} = (4\pi e^2 /\hbar) N_{L/R}(0) N_I(0) |T_{L/R}|^2 $,
where $N_r(0)$ are the densities of states of the
the electrodes and the island, $r = L,R,I$.
We further assume that the junctions, although small,
still accommodate many transverse channels. From a comparison
of Andreev reflection and single electron tunneling in small
normal-superconducting junctions we have
concluded that the number of channels is $N_{ch} \ge 10^3$.

The quantum mechanical many-body problem of electrons coupled by Coulomb
interactions can be reformulated in a path integral representation. In order
to handle the interaction one performs a Hubbard-Stratonovich transformation.
It introduces collective  variables $\varphi_r$, which are the quantum
mechanical conjugates of the charges in the electrodes or island.
The capacitive interaction between electrons
is replaced in this way by an interaction of the electrons with the fields
$V_r \equiv \hbar \dot{\varphi}_r/e$.
After this stage the electronic degrees of freedom can be traced out.
The next step  involves an expansion in the electron propagators.
Since we consider wide junctions only simple loops need to be retained
\cite{bruder}. Their iteration introduces in each order a factor  $N_{ch}$,
hence they dominate over more complicated higher order loops.
After the trace has been performed the system can no longer be described
by a Hamiltonian. Instead we deal with a reduced density matrix
$\rho(\{\varphi_{r,1}\},\{\varphi_{r,2}\})$. It
depends on an effective action, which is expressed in the phases
$\varphi_{r,\sigma}$
corresponding to the forward and backward propagator $\sigma =1,2$.
The structure of the theory is familiar from Refs.
 \cite{Fey-Ver,CLeggett}, where a quantum system coupled to a harmonic
 oscillator bath has been considered. The analogous model
describing the effect of electron tunneling has been presented in
Refs. \cite{Eck-Scho-Amb,Schoen-Zai}.
We, therefore, do not present the derivation of the steps here, rather we
quote only the result.

For transparency we describe the formalism for a single junction.
The time evolution of its reduced density matrix is given by
($\hbar=k=1$)
\begin{eqnarray}
	\rho(t_f;\varphi_{1f},\varphi_{2f})=
	\int d\varphi_{1i} d\varphi_{2i}
	\int_{\varphi_{1i}}^{\varphi_{1f}} \cal{D}\varphi_1(t)
	\int_{\varphi_{2i}}^{\varphi_{2f}} \cal{D}\varphi_2(t)
\nonumber\\
	e^{iS[\varphi_1,\varphi_2]}
	\rho(t_i;\varphi_{1i},\varphi_{2i}) \, .
\end{eqnarray}
Here $\varphi_1$ and $\varphi_2$ refer to the forward and backward time
evolution. The effective action is given by
\begin{equation}
        S[\varphi_1,\varphi_2] = S_{ch}[\varphi_1] - S_{ch}[\varphi_2]
                                + S_t[\varphi_1,\varphi_2] \; ,
\end{equation}
The first term represents the charging energy $ S_{ch}[\varphi]
= \int_{t_i}^{t_f} dt \frac{C}{2}
                        \left(\dot{\varphi}/ e\right)^2 $.
Electron tunneling is described by \cite{Eck-Scho-Amb,Schoen-Zai}
\begin{equation}
	S_t[\varphi_1,\varphi_2] = 4 \pi i \sum_{\sigma,\sigma'=1,2}
	\int_{t_i}^{t_f} dt
	\int_{t_i}^{t} dt' \alpha^{\sigma,\sigma'}(t-t')
	\cos[\varphi_{\sigma}(t) - \varphi_{\sigma'}(t')] \; ,
\end{equation}
where $\alpha^{\sigma,\sigma'}$ are given in Fourier space by
\begin{eqnarray}
     	\alpha^{\sigma,1}(\omega) = (-1)^{\sigma+1} \alpha^- (\omega)
	\qquad , \qquad \alpha^{\sigma,2}(\omega) = (-1)^{\sigma}
	\alpha^+ (\omega)\\
	\mbox{and} \qquad \alpha^{\pm}(\omega)=\pm\alpha_0\,\omega \,
	\frac{1}{\exp(\pm\beta \omega)-1}  \; .
\label{25}
\end{eqnarray}
The tunneling term couples the forward and backward propagators.
This arises in the step where the microscopic degrees of freedom
are eliminated.

An important step for a systematic description of tunneling
processes is the change from the phase
 to a charge representation, accomplished by
\begin{eqnarray}
        \rho(t_f;n_{1f},n_{2f}) = \sum_{n_i}\rho(t_i;n_i,n_i)
        \int d\varphi_{1f} d\varphi_{2f} d\varphi_{1i} d\varphi_{2i}\nonumber\\
	\int_{\varphi_{1i}}^{\varphi_{1f}} \cal{D}\varphi_1(t)
        \int_{\varphi_{2i}}^{\varphi_{2f}} \cal{D}\varphi_2(t)
	\int \cal{D} n_1(t) \int \cal{D} n_2(t) \nonumber\\
	\times \exp\left(-in_i\varphi_{1i}+in_{1f}\varphi_{1f}-iS_{ch}[n_1]
	+i\int d t \, n_1 \dot{\varphi}_1\right)\nonumber\\
	\times \exp\left(+in_i\varphi_{2i}-in_{2f}\varphi_{2f}+iS_{ch}[n_2]
	-i\int d t \, n_2 \dot{\varphi}_2\right)\nonumber\\
	\times \exp{\{4 \pi i \sum_{\sigma,\sigma'=1,2}
	\int_{t_i}^{t_f} dt
	\int_{t_i}^{t} dt' \alpha^{\sigma,\sigma'}(t-t')
	\cos[\varphi_{\sigma}(t) - \varphi_{\sigma'}(t')]\}} \; .
\end{eqnarray}
Here $S_{ch}[n] = \int_{t_i}^{t_f} dt \frac{1}{2C}(ne)^2 $.
In systems with discrete charges the integrations over $\varphi$
include a summation over winding numbers \cite{Schoen-Zai}.
We further assume that the initial density matrix
is diagonal in the charges.
In the absence of tunneling the bare forward and backward propagators
involve only the charging energy $\exp(\pm i S_{ch}[n])$.
We expand the tunneling part of the action $\exp(iS_t)$ and
integrate over $\varphi$.
Each of the exponentials $\exp[\pm i\varphi_{\sigma}(t)]$
causes a change of the number of electrons on the island by $\pm 1$
at time $t$ on the forward or backward branch, $\sigma = 1$ or $2$,
respectively.
These changes occur in pairs and are connected by a line representing
$\alpha^{\sigma,\sigma'}(t-t')$. The two correlated transitions can
occur on the same branch $\sigma = \sigma'$, or on different branches.

The latter are of particular interest.
Imagine we started in a state with n charges
$\rho(t_i) = |n\rangle \langle n|$.
Then each transition, described by $\exp(i[\varphi_1(t)-\varphi_2(t')])$
changes the charge on both branches by $+e$ and the density matrix acquires a
finite value also for states $|(n+1)e \rangle$.
After integrating over the two times $t$ and $t'$, limited by
$t_i \le t' \le t \le t_f$,  we find
\begin{equation}
	\langle n+1 | \rho(t_f) | n+1 \rangle
	= (t_f-t_i) 2 \pi \alpha^+(\Delta E_{ch}(n)) \; ,
\end{equation}
where $\Delta E_{ch}(n) = E_{ch}(n+1)-E_{ch}(n)$.
Obviously we can interpret the coefficient of the time difference as
transition rate, and indeed we reproduce the well-known single electron
tunneling rate.

In the following we consider the single electron transistor.
Each capacity $C_r$ introduces a charging term $S_{ch}$,
where $r=L,R,G$ denotes the reservoirs on the left, right and the gate,
and each tunnel junction introduces a term $S_t$, all depending on the
appropriate phase difference $\varphi_I - \varphi_r$.
The phases in the reservoirs $r$ are assumed to be
well-defined quantities $\varphi_r=eV_r t$ without fluctuations.
The contribution of the charging energy which is related to the work done by
the transport voltage can be accounted for in the tunneling lines by
$\alpha^{\pm}_r(\omega-eV_r)$, and the charging energy becomes
$E_{ch}(n)= (ne-{Q_G})^2/(2C)$.

Each term of the expansion can be visualized by a diagram.
In Fig. (\ref{fig2}) we show several important processes.
\begin{figure}
\vspace{2.7cm}
\caption{
Example of a diagram showing various tunneling processes:
on the left sequential tunneling in the left and right junctions,
then a term which preserves the norm, next a cotunneling process,
and on the right resonant tunneling in the left junction.
}
\label{fig2}
\end{figure}
There is a closed time-path consisting of two horizontal lines joined at $t_f$.
They correspond to the forward propagator from $t_i$
to $t_f$ (upper line) and the
backward propagator from $t_f$ to $t_i$ (lower line).
Along the time-path we arrange vertices. They are connected in pairs
by (dashed) tunneling lines,
either within one propagator or between the two propagators.
We draw all topological different diagrams with
directed tunneling line and evaluate the diagrams
according to the following rules:
\begin{description}
\item[  1.  ]
Assign charge states $n$ and the corresponding charging energy
to each element of the propagators. Elements of the forward (backward)
propagator between $t$ and $t'<t$ carry factors
$\exp[\mp i E_{ch}(n)(t-t')]$.
\item[  2.  ]
Each vertex represents an exponential $\exp[\pm i \varphi_{\sigma}(t)]$
of the tunneling contribution to the action. It changes the charge
from $n$ to $n\pm1$.
\item[  3.  ]
Pairs of vertices are connected by a directed tunneling line
$\alpha^+_r(t-t') \, [\alpha^-_r(t-t')]$ for the electrodes $r=L,R$,
if the line of
is running backward [forward] with respect to the closed time-path.
The charge increases (decreases) along the time-path
by $1$ if a tunneling line comes in (goes out).
\item[  4.  ]
Each diagram carries an prefactor $(-i)^M(-1)^m$, where M is the total
number of vertices and m their number on the
backward propagator.
\item[  5.  ]
Integrate over the internal times and sum over the reservoirs.
\end{description}

In order to calculate stationary transport properties it is convenient to
change to an energy representation.
This is achieved by the following transformation.
In each diagram we order the times from left to right and label them by
$t_j$, irrespective on which branch they are.
We further set $t_i =- \infty$ and $t_f = 0$.
We then encounter integrals of the type
$$
	\int^0_{-\infty} dt_1 \int^0_{t_1} dt_2
	\ldots \int^0_{t_{M-1}} dt_M \, e^{\eta t_1}
	e^{-i \Delta E_1(t_2-t_1)}
	e^{-i \Delta E_2(t_3-t_2)} \cdots
	e^{-i \Delta E_M(-t_M)}
$$
$$
	=\,(-i)^M \frac{1}{\Delta E_1-i\eta}\, \cdot \,
	\frac{1}{\Delta E_2-i\eta}
	\cdots \frac{1}{\Delta E_M -i\eta} \; .
$$
Here $\Delta E_j$ is the difference of the energies of the upper and
lower propagator and -- if present -- the frequency of
the tunneling line within the segment limited by $t_{j}$ and $t_{j+1}$.
The convergence factor $e^{\eta t_1}$ ($\eta\rightarrow 0^+$) is
related to an adiabatic switching of the tunneling term $H_t$. The
rules in energy representation read:
\begin{description}
\item[  1.  ]
Draw all topological different diagrams.  These are the same as in time space.
In addition to the charging energy assigned to the propagators
we assign a frequency $\omega$ to each tunneling line.
\item[  2.  ]
For each segment derived from $t_{j} \le t \le t_{j+1}$ with $j \ge1$
we assign a resolvent ${1\over \Delta E_j  -i\eta}$ where
$\Delta E_j$ is the difference of the energies
of the forward and backward propagator,
plus the sum of the frequencies of the tunneling lines in the given segment.
The latter have to be taken positive for lines from the left to the right and
negative for lines from the right to the left.
\item[  3.  ]
The prefactor is given by $(-1)^{m+l}$, where m is the total number
of vertices on the backward propagator and $l$ the total number of resolvents.
\item[  4.  ]
For each coupling of vertices we write $\alpha^+_r(\omega) \,
[\alpha^-_r(\omega)]$, if the tunneling line of reservoir $r$ is going
backward (forward) with respect to the closed time-path.
\item[  5.  ]
Integrate over the frequencies of tunneling lines and sum over the reservoirs.
\end{description}

\section{Master equation and stationary probabilities}

We denote the sum of all diagrams by
$\Pi^{n_1,n'_1}_{n_2,n'_2}$, where the indices
indicate the left and right charge states on the two branches,
$\sigma = 1,2$.
They can  be expressed by  an irreducible self-energy part
$\Sigma^{n_1,n'_1}_{n_2,n'_2}$,  defined as
the sum of all diagrams where any vertical line cutting through them
crosses at least one tunneling line.
The time evolution of the density matrix can be expressed as an iteration in
the style of a Dyson equation, illustrated in Fig.(\ref{fig3})
\begin{equation}
	\Pi^{n_1,n'_1}_{n_2,n'_2} =
	{\Pi^{(0)}}^{n_1}_{n_2} \delta_{n_1,n_1'} \delta_{n_2,n_2'} +
	\sum_{n''_1,n''_2}
	\Pi^{n_1,n''_1}_{n_2,n''_2}\, \Sigma^{n''_1,n'_1}_{n''_2,n'_2} \,
	{\Pi^{(0)}}^{n'_1}_{n'_2} \; .
\label{pi}
\end{equation}
Here $\Pi^{(0)}$ is the free propagator.

\begin{figure}[b]
\vspace{2.5cm}
\caption{
The iteration of processes for $\Pi$, describing the time evolution
of the density matrix.
}
\label{fig3}
\end{figure}

We start from an density matrix which is diagonal,
 $\rho(t_i,n_1,n_2) =\\
 P^{(0)}(n_1) \delta_{n_1,n_2}$.
This means that the density matrix remains diagonal except during
transitions described by $\Sigma$. Hence, we consider
$\Sigma_{n,n'} \equiv \Sigma^{n,n'}_{n,n'} \;.$
We identify the solution of Eq. (\ref{pi}) -- after multiplication with
$P^{(0)}(n)$ from the left -- as the stationary distribution function
$\sum_n P^{(0)}(n)\Pi^{n,n'}_{n,n'} = P^{st}(n')$.
Hence Eq. (\ref{pi}) reduces to
\begin{equation}
	P^{st}(n') = P^{(0)}(n') +  \sum_{n''}
	P^{st}(n'') \Sigma_{n'',n'} \Pi^{(0)}_{n',n'}\; .
\label{pi0}
\end{equation}
The last term in Eq. (\ref{pi0}), $\Pi^{(0)}$, describes a propagation
in a diagonal state (i.e. equal energies on the forward and backward
propagator) which does not contain a tunneling line.
According to the rules in the energy representation this introduces a factor
$1/i\eta$, and in the limit $\eta \rightarrow 0+$ we find
$\sum_{n'}P^{st}(n') \Sigma_{n',n} =0\;.$\label{pi00}
By attaching the rightmost vertex of each diagram $\Sigma$
to the upper and lower propagator we can show that
\label{30}
$\sum_{n^\prime}\Sigma_{n,n^\prime}=0 \,$ ,
which allows us to rewrite Eq.(\ref{pi00}) also in the form
\begin{equation}\label{31}
	0 = - P^{st}(n) \sum_{n^\prime \ne n}\Sigma_{n,n^\prime}
	+ \sum_{n^\prime \ne n}P^{st}(n^\prime)\Sigma_{n^\prime,n} \;.
\end{equation}
We recover the structure of a stationary master equation
with transition rates given by $\Sigma_{n',n}$.
In general the irreducible self-energy $\Sigma$ yields
the rate of all possible correlated tunneling processes.
We, furthermore, see that the stationary solution $P^{st}(n)$
is independent of the initial distribution $P^{(0)}(n)$.

For illustration we evaluate now all diagrams which contain no overlapping
tunneling lines. After each transition the system is in a diagonal state of the
density matrix and a probability can be assigned. Examples are
 visualized on the left hand side of Fig. (\ref{fig2}).
This means we include those processes which are also described
by the master equation with rates obtained in lowest order perturbation
theory.
In the limit considered the irreducible part $\Sigma^{(1)}$ consists
of two classes of diagrams, those where a tunneling line connects two
vertices on one propagator (which implies that $n'_i = n_i$)
and those where it connects the two propagators ($n'_i = n_i \pm 1$, with
$n_1-n_2=n'_1-n'_2$) (see Fig.(\ref{fig4})).
According to the rules the rates are
\begin{equation}
	\Sigma^{(1)}_{n,n\pm1} = 2\pi i\alpha^\pm (\pm \Delta E_{ch}^\pm)
	\qquad , \qquad
	\Sigma^{(1)}_{n,n} = -2\pi i \sum_\pm\alpha^\pm
	(\pm \Delta E_{ch}^\pm)
\label{sigma}
\end{equation}
where $\Delta E_{ch}^\pm =E_{ch}(n\pm1)-E_{ch}(n)$.
\begin{figure}[h]
\vspace{2.7cm}
\caption{
Representation of the self-energy $\Sigma^{(1)}$,
defined to contain no overlapping tunneling lines.
Only one representative of each class is shown,
the remaining ones are obtained by changing the direction of the arrows
and exchanging the position on the forward and backward propagator.
}
\label{fig4}
\end{figure}

\section{Higher order tunneling processes}

\subsection{Cotunneling}

In situations where single electron tunneling is suppressed
by Coulomb blockade the lowest order contribution to the current arises
due to cotunneling. It is described by a diagram
with tunneling processes in the left and in the right junction, where
the corresponding lines $\alpha_L(t_L-t_L')$ and $\alpha_R(t_R-t_R')$
overlap in time. This means there is no intermediate time
where the density matrix is diagonal, which would describe
sequential tunneling.  An example is shown in Fig. (\ref{fig2}).
Performing the integrations we find for the rate
\begin{eqnarray}
	\Sigma_{cot} = \frac{\hbar}{2\pi e^4R_L R_R}
	\int d\epsilon_1 d\epsilon_2 d\epsilon_3 d\epsilon_4
	f(\epsilon_1)[1-f(\epsilon_2)]
	f(\epsilon_3)[1-f(\epsilon_4)]\nonumber\\
	\times \left (\frac{1}{\epsilon_2-\epsilon_1+E_1}
	+\frac{1}{\epsilon_4-\epsilon_3+E_2}\right)^2
	\delta(eV+\epsilon_1-\epsilon_2+\epsilon_3-\epsilon_4) \; ,
\end{eqnarray}
plus a divergent term.
This divergence is canceled if we take into account consistently all
processes of the same order. This includes a process
where the island has already the charge $(n\pm1)e$ and one tunneling event
brings it back to $ne$.
The corresponding terms are products of a single electron  tunneling rate of
one junction and the stationary probability $P^{st}_{n\pm1}$ in first
order in $\alpha_0$.
Adding all these terms we obtain the result given above.

\subsection{Resonant Tunneling}

The perturbative approach breaks down at low temperatures or for large
values for the dimensionless conductance $\alpha_0$. Specifically we
will show that the classical master equation is valid only for
$\alpha_0 \ln{({E_C\over 2\pi T})}\ll 1$, whereas for larger
values resonant tunneling processes become important.

To proceed we have to find a systematic criterion which diagrams should be
retained and summed. For this we
note that during a tunneling process the reservoirs contain an electron
excitation. Two parallel tunneling lines imply two such excitations.
Our criterion is that we take into account only those matrix elements of the
total density matrix, i.e. reservoirs plus charge states, which differ at most
by two excitations in the leads or (equivalently) in
the island. Graphically this means that any vertical line will cut at most
two tunneling lines.

We will concentrate here on situations where only two charge states
with $n=0, 1$ need to be considered. This is sufficient when the
energy difference of the two charge states
$\Delta_0 \equiv E_{ch}(1)-E_{ch}(0)$ and
the bias voltage $eV=eV_L-eV_R$
are small compared to $E_C$, which is the energy
associated with transitions to higher states.
The combination of the two restrictions, diagrams with
at most two parallel tunneling lines in the two-state problem,
implies that no diagram contains $crossing$ tunneling lines.
In this case we can evaluate the irreducible self-energy analytically.

In order to sum all diagrams which may contain up to
two parallel tunneling lines we introduce a diagram labeled by
$\phi_{n_2,n_2'}^{n_1,n_1'}(r,\omega)$
(see Fig.(\ref{fig5})).
\begin{figure}[h]
\vspace{5.0cm}
\caption{
a) Self-consistent equation for $\phi^r_n(\omega)$. A summation over
the electrodes $r$ and the direction of the tunneling lines is implied.
b) Representation of the self-energy $\Sigma_{n,1}$ within our
approximation.
}
\label{fig5}
\end{figure}
It has an open tunneling line associated with
tunneling in the junction $r$ carrying the frequency $\omega$.
Consequently it remains in an off-diagonal state at
one side. For the two state problem we need only
\begin{equation}
	\phi_n^r(\omega) \equiv  \phi_{n,0}^{n,1}(r,\omega)
\label{phi}
\end{equation}
with $n=0,1$, for which we can formulate the
iteration depicted diagrammatically in Fig.(\ref{fig5}). It yields
\begin{equation}
	\phi^r_n (\omega)
	=\pi(\omega)\left[\alpha^+_r(\omega)\delta_{n,0}
	-\alpha^-_r(\omega)\delta_{n,1}+
	\alpha_r(\omega)\int d\omega^\prime\,
	\sum_{r^\prime}
	\frac{\phi^{r^\prime}_n(\omega^\prime)^*}{\omega^\prime-\omega-i\eta}
	\right].
\label{34}
\end{equation}
Here we encounter the propagator
$\pi(\omega) \equiv {\Pi^{(1)}}_{0,0}^{1,1}(\omega)$.
It is given by the propagator $\Pi$,
restricted to first order in the tunneling,
and we start and end in an off-diagonal state.
Furthermore, since the parallel tunneling line carries a frequency $\omega$
the energies of the upper and lower lines are effectively shifted
relative to one another.
Notice that due to the restriction to a two state problem
there are no diagrams contained in ${\Pi^{(1)}}_{0,0}^{1,1}$
where a tunneling line connects the upper and lower propagator.
We can express it by the  first order self-energy
$\sigma(\omega) \equiv {\Sigma^{(1)}}_{0,0}^{1,1}(\omega)$,
which is the off-diagonal version of the expression
known from the first order discussion, with the
added complication of a parallel tunneling line with frequency $\omega$.

The irreducible self-energy $\Sigma$ is obtained from $\phi(\omega)$
by connecting the tunneling line with the appropriate direction to the
upper and the lower propagator and adding both contributions
(see Fig.(\ref{fig5})). We get for $n=0,1$
\begin{equation}\label{33}
	\Sigma_{n,1} = -2i\,Im \int d\omega\sum_r\phi^r_n(\omega) \; ,
\end{equation}
while $\Sigma_{n,0}$ follows from the rule $\Sigma_{n,0}+\Sigma_{n,1}=0$.

Applying our rules for the calculation of the diagrams we find
\begin{equation}\label{35}
	\pi(\omega)=
	{1\over\omega-\Delta_0-\sigma(\omega)}\qquad,\qquad
	\sigma(\omega)= -\int d\omega^\prime{\alpha(\omega^\prime)
	\over \omega^\prime-\omega-i\eta}.
\end{equation}
Here and for the following we introduce the notations
$\alpha(\omega)=\alpha^+(\omega)+\alpha^-(\omega)$,
$\alpha^\pm(\omega)=\sum_r\alpha^\pm_r(\omega)$,
$\alpha_r(\omega)=\alpha^+_r(\omega)+\alpha^-_r(\omega)$
and $\alpha_0=\sum_r\alpha_0^r$ .
Notice that
$\alpha^\pm_r(\omega)=\alpha_r(\omega)f(\pm(\omega-eV_r))$ where
$f(\omega-eV_r)=f_r(\omega)=1/[\exp(\beta (\omega-eV_r))+1]$ is the Fermi
function of reservoir $r$.

We solve the integral equations (\ref{34}) and obtain
\begin{eqnarray}
\Sigma_{0,1}=-\Sigma_{0,0}=2\pi i{\lambda_+\over\lambda}\qquad , \qquad
\Sigma_{1,0}=-\Sigma_{1,1}=2\pi i{\lambda_-\over\lambda}\label{40}\\
\mbox{with}\qquad
\label{39}
\lambda_\pm=\int d\omega\alpha^\pm(\omega)|\pi(\omega)|^2\qquad,\qquad
\lambda=\int d\omega |\pi(\omega)|^2.
\end{eqnarray}
Inserting these quantities in the kinetic equation (\ref{31})
we arrive at the stationary probabilities
$P_0^{st}=\lambda_-$ and $P_1^{st}=\lambda_+\,$.
Both probabilities are positive and normalized $\lambda_+ +\lambda_- =1$.

\subsection{The Current}

The expression for the current at time $t$ in the junction $r$ can
be written as
\begin{equation}
	I_r (t) = -2ie \langle \int^{t}_{-\infty} dt' \sum_{\sigma}
	\alpha^{1,\sigma}_r(t-t')
	\sin[\varphi_{r,1}(t)-\varphi_{r,\sigma}(t')]
	\rangle \, ,
\label{current}
\end{equation}
where the expectation value is taken with the reduced density matrix
discussed above, where the larger of the times is $t=t_f$.
If we express the expectation value in the charge representation
we see that the sin-functions in Eq.(\ref{current}) describe the
transfer of charges at times $t_f$ and $t'$. We, therefore, have to
evaluate the two real-time correlation functions
describing charge transfer processes at times $t$ and $t^\prime$
\begin{equation}
	C^>(t,t^\prime)=-i{\langle e^{-i\varphi(t)}e^{i\varphi(t^\prime)}
	\rangle} \qquad , \qquad
	C^<(t,t^\prime)=i{\langle e^{i\varphi(t^\prime)}e^{-i\varphi(t)}
	\rangle} \; .
\label{13}
\end{equation}
In the stationary limit the current can be expressed by
\begin{equation}
	I^{st}_r=-ie\int d\omega\left\{\alpha^+_r(\omega)C^>(\omega)+
	\alpha^-_r(\omega)C^<(\omega)\right\}.
\label{42}
\end{equation}
It is further convenient to introduce a spectral density for charge excitations
\begin{equation}
	A(\omega)={1\over 2\pi i}[C^<(\omega)-C^>(\omega)] \; .
\label{15}
\end{equation}

The correlation functions and spectral density can be evaluated in the
approximation which we have used before,
\begin{eqnarray}
        C^{<(>)}(\omega)&=&+(-)2\pi i \sum_r\alpha_r(\omega)
	f(+(-)[\omega-eV_r])
	|\pi(\omega)|^2 \label{46d}\\
	\mbox{and}
	\qquad A(\omega)&=&\alpha(\omega)\,|\pi(\omega)|^2\label{45} \; .
\end{eqnarray}
The current can then be written as
\begin{equation}
	I^{st}_r={e\over h}4\pi^2\int d\omega\sum_{r^\prime}
	{\alpha_{r^\prime}(\omega)\alpha_r(\omega)\over\alpha(\omega)}
	A(\omega) [f_{r^\prime}(\omega)-f_r(\omega)] \; .
\label{46c}
\end{equation}

These results satisfy the conservation laws and sum
rules. The current is conserved, $\sum_r I^{st}_r=0$, and vanishes in
equilibrium when $V_r=0$. The spectral density is normalized
 $\int d\omega A(\omega) =1$, and the usual relationships
between the correlation functions and the spectral density in
equilibrium are reproduced. The classical result can be recovered
if we use the lowest order approximation in $\alpha_0$ for the
spectral density
$A^{(0)}(\omega)=\delta(\omega-\Delta_0)$.
We conclude with the observation that quantum
fluctuations yield energy renormalization and broadening effects,
which manifest themselves in the spectral density via  the real and
imaginary part of the self-energy $\sigma(\omega)$ given in
Eq.(\ref{35}). It will be evaluated in the next section.

\section{Results and applications}

\subsection{Charge Fluctuations in the Single Electron Box}

In equilibrium when $V_R=V_L$, the SET-transistor is
equivalent to the single electron box. The average excess particle
number can be obtained from (\ref{39}) and (\ref{45})
$\langle n \rangle=\int d\omega f(\omega) A(\omega)$.
In the classical limit
$A^{(0)}(\omega)=\delta(\omega-\Delta_0)$ one obtains
$\langle n^{cl} \rangle =f(\Delta_0)$
where the energy difference $\Delta_0=E_C(1-2C_GV_G)$ depends on
the gate voltage. Thus, $\langle n^{cl}(Q_G) \rangle$ shows a
step at $Q_G = C_G V_G = 1/2$, which is smeared by
temperature.

At larger values of $\alpha_0$ or lower temperature we have to include
the self-energy $\sigma(\omega)$ (\ref{35}) in the spectral density
(\ref{45}). The two limits, $T=0$ and $|\omega|\le T$, can be
analyzed analytically.
In the first case, the spectral density has the form
\begin{equation}\label{52}
	A(\omega)\cong{|\omega|\over\Delta_0}\,\cdot\,
	\frac{
	\tilde{\Delta}(\omega) \tilde{\alpha}(\omega)}
	{[\omega-\tilde{\Delta}(\omega)]^2
	+[\pi \tilde{\Delta}(\omega) \tilde{\alpha}(\omega)]^2}\,\,,\qquad T=0
\end{equation}
where
\begin{equation}
	\tilde{\Delta}(\omega)=\frac{\Delta_0}{1+2\alpha_0
	\ln({E_C\over|\omega|})}
	\frac{1}{1+\pi^2\tilde{\alpha}(\omega)^2}\; \;  ,\; \;
	\tilde{\alpha}(\omega)
	={\alpha_0\over 1+2\alpha_0 \ln({E_C\over
	|\omega|})} \; .
\label{54}
\end{equation}
The spectral density $A(\omega)$ has a maximum at the renormalized
energy difference $\Delta$, which is obtained from
the self-consistent solution of
\begin{equation}
	\Delta=\tilde{\Delta}(\Delta) \; .
\label{delta}
\end{equation}
It further has a broadening of order $\pi\Delta\alpha$
 where $\alpha=\tilde{\alpha}(\Delta)$.
For $\pi\alpha\ll 1$ the broadening can be
neglected. In this case our results coincide with the RG-analysis of Refs.
\cite{Fal-Scho-Zim,Mat}, which shows that the leading logarithmic terms are
included in our diagram series.

At finite temperatures $|\omega|\le T$, we get
\begin{equation}\label{56}
	A(\omega)\cong{\Delta\over
	\Delta_0}\,\cdot\,{\alpha\,\omega\,
	\coth{({\omega\over 2T})}\over
	[\omega-\Delta]^2+[\pi\alpha\,\omega\,
	\coth({\omega\over 2T})]^2}\quad,\quad|\omega|\le T
\end{equation}
where
\begin{equation}
	\Delta =\frac{\Delta_0}{1+2\alpha_0 \ln({E_C\over 2\pi T})}
	\; \;  ,\; \;
	\alpha={\alpha_0\over 1+2\alpha_0 \ln({E_C\over
	2\pi T})} \; .
\label{56b}
\end{equation}
The broadening is of order $\pi\alpha T$. It can only be neglected if
for $\pi\alpha \ll 1$.

As a consequence of quantum fluctuations the step of
the average charge in the electron box near the degeneracy points
is washed out.
We neglect broadening effects ($\pi\alpha\ll1$) and assume that
the energies of the ground state and the
first excited state near the degeneracy point depend symmetrically
on the distance,
${E_C\over4}\mp{\Delta\over 2}$.
In this case the partition function is $Z\cong\,2\,e^{-{E_C\over 4T}}
\,\cosh({\Delta\over 2T})$ and  the average excess charge,
$\langle n \rangle=Q_G/e-T{\partial\over\partial \Delta_0}\ln Z$,
becomes
\begin{equation}\label{63}
	\langle n \rangle \cong
	{\Delta_0-\Delta \over 2\Delta_0}+{\Delta
	\over \Delta_0}f(\Delta) \, .
\end{equation}
Depending on the temperature we have to choose the appropriate limiting form
for $\Delta$. At finite temperature the slope at  $\Delta_0=0$ is
$
\partial \langle n \rangle /\partial\Delta_0|_{\Delta_0=0}
=-\{4T[1+2\alpha_0 \ln{({E_C\over 2\pi T})}]^2\}^{-1}$. It is
reduced compared to the classical result $-1/4T$.
This anomalous temperature dependence
can serve as an indication for coherent higher order tunneling processes.
Below we will demonstrate that the linear and nonlinear conductance of a
transistor shows more pronounced deviations from the classical result,
making it more suitable for experimental verification.

\subsection{Conductance oscillations in the SET-transistor}

The linear conductance G of a transistor follows from (\ref{46c})
\begin{equation}\label{66}
	G=-{e^2\over h} 4\pi^2\int d\omega \;
	{\alpha_R(\omega)\alpha_L(\omega)\over\alpha(\omega)}
	A(\omega)f^\prime(\omega).
\end{equation}
Since the derivative of the Fermi function restricts the integration variable
to the regime $|\omega|\le T$  we can use the form (\ref{56})
and obtain
\begin{equation}\label{67}
	G={e^2\over h}2\pi^2\,\,{\alpha_R\alpha_L\over\alpha}
	\int d\omega \; {\omega/T\over\sinh(\omega/T)}\,\cdot\,
	{\alpha\,\omega\, \coth{({\omega\over2T})}
	\over[\omega-\Delta]^2
	+[\pi\alpha\,\omega\, \coth{({\omega\over 2T})}]^2}
\end{equation}
where  $\alpha=\alpha_R+\alpha_L$ and we have to use
the finite temperature form for $\Delta$ and $\alpha$.
The conductance has a maximum at $\Delta_0=0$, given by
\begin{equation}\label{68}
G_{max}={e^2\over h} 2\pi{\alpha_R\alpha_L\over\alpha}\left[{\pi\over
2}-\arctan \left({(\pi\alpha)^2-1\over 2\alpha\pi}\right)\right] \, ,
\end{equation}
with a broadening given by
\begin{equation}\label{70}
	\gamma\cong{\int d\Delta_0 \,
	G(\Delta_0)\over G_{max}}
	={\pi^3\,T[1+2\alpha_0 \ln{({E_C\over 2\pi T})}]\over
	\pi-2\,\arctan[{(\pi\alpha)^2-1\over2\alpha\pi}]}.
\end{equation}
In the regime $2\alpha_0 \ln{({E_C\over 2\pi T})}\ll1$ the peak hight $G_{max}$
is constant and $\gamma$ proportional to $T$, which is the classical result
for sequential tunneling. For $2\alpha_0 \ln{({E_C\over 2\pi T})}\sim 1$
and $\pi\alpha\ll 1$,
$G_{max}$ and $\gamma$ contain logarithmic terms in the temperature
indicating energy renormalization effects due to
higher-order tunneling processes. For $T\rightarrow 0$
 the maximum value as well as the broadening vanish,
$G_{max}\sim{1\over \ln T}$ and $ \gamma\sim T\ln T$.
For $2\alpha_0 \ln{({E_C\over 2\pi T})}\sim 1$ and $\pi\alpha\sim 1$,
we have to retain the finite life-time
effects contained in (\ref{68}) and (\ref{70}).

\begin{figure}
\vspace{9.5cm}
\caption{
The differential conductance in the nonlinear response regime as function
of the gap energy normalized to the transport voltage $V$,
for $T=0$ and $\alpha_0^{L}= \alpha_0^{R}=0.05 \;(R_L=R_R=R_T)$,
we consider a symmetric bias and chose $(1)\,V/E_C=0.1\,,\,(2)\,V/E_C=0.01\,,\,
(3)\,V/E_C=0.001\,$. For comparison, $(0)$ shows the result for the classical
case obtained from lowest order perturbation theory and the master equation.
}
\label{fig7}
\end{figure}
The most pronounced signature of quantum fluctuations is contained in the
differential conductance $G(V)=\frac{\partial{I^{st}(V)}}{\partial{V}}$.
In this case the finite voltage $eV$ provides another energy scale
and the renormalization and life-time effects are probed over
a wider energy range even at zero temperature.
The $T=0$ result obtained from Eq. (\ref{46c}) is plotted in Fig.(\ref{fig7}).
For comparison we also show the result obtained in perturbation theory.
(We had quoted the expression for the current in this limit in the
Introduction.)
In the classical limit the conductance  is nonzero only in the range
$|\Delta_0|\le\frac{eV}{2}$ with vertical steps at the edges.
The result of Fig.(\ref{fig7}) displays clearly the renormalization effects
and, moreover,  the finite life-time broadening. It is clear
that at finite temperature the conductance is washed out further.
These effects should be
observable in an experiment with realistic parameters.

\section{Interactions in the leads}

We will now address the problem of interactions in the leads
by modeling the leads as 1-dimensional Luttinger liquids.
In contrast to Kane and Fisher \cite{Kane2} we assume
that the interactions in the island are sufficiently described by the
charging energy. We derive a path integral description for this case
and find modifications of the effective action as compared to the
metallic case. However,
the physics associated with the Coulomb blockade remains the same.

The Hamiltonian of the system is still given by Eq. (\ref{ham}), where
the Hamiltonians of the leads describe Luttinger liquids. They are
given by~\cite{Kane}
\begin{equation}
	H_L =
	\int \frac{dx}{2 \pi } \sum _{j} v_j
	\left[	\frac{g_j}{2} (\nabla \phi _j )^2 +
		\frac{1}{2g_j} (\nabla \theta  _j)^2
	\right] \, ,
\label{lutham}
\end{equation}
i.e. they contain a sum of spin ($j = \sigma $) and charge ($j = \rho $)
degrees of freedom. The parameters $g_j$ describe the
interaction strength ($g_j = 2$ in the noninteracting limit),
and $v_j$ are the velocities  of spin and charge excitations.
We also introduce the bosonic fields
$\phi _{s}= \phi _{\rho} +s \phi _{\sigma}$
and
$\theta _{s}= (1/2)[\theta _{\rho} +s \theta_{\sigma}]$
for spin up ($s=1$) and down ($s=-1$) fermions.
These fields obey the commutation relation
$[\phi _s (x), \theta _{s'} (x')] =
(i \pi ) \Theta (x-x') \delta _{s,s'}$.

The fermionic field operators $\hat{\Psi}$
can be expressed in terms of the spin and charge degrees of
freedom ~\cite{Haldane2}
\begin{equation}
	\hat{\Psi} ^{\dagger}_{L,s}(x,\tau)
	=
	\sqrt{\rho _{0,s}}
	\sum \limits _{m=\pm 1}
	e^{i m k_F x }
	e^{i \frac{m}{2} [\theta _{\rho} +s \theta _{\sigma}]}
	e^{i [\phi _{\rho} +s \phi _{\sigma}]},
\label{fieldoperator}
\end{equation}
where $\rho _{0,s} = N_{s}/2L$ is the electron density for one spin
direction and $k_F$ the Fermi wave-vector.
The bosons fields $\theta $ and $\phi $ can be decomposed in terms of scalar
bosonic fields  $\hat{b}_{j,q}, \hat{b}^{\dagger}_{j,q}$ ~\cite{Haldane2}
\begin{eqnarray}
	\theta _j(x;\tau)
	&=&
	i \sqrt{g_j} \sum _{q \ne 0}
	\left|
	\frac{\pi}{2qL}
	\right|^{1/2}
	\mbox{sign}(q) e^{iqx}
	(e^{|q|v_{j} \tau} \hat{b}^{\dagger}_{j,q}
	+
	e^{-|q|v_{j} \tau} \hat{b}_{j,-q}) \nonumber \\
	\phi _j(x;\tau)
	&=&
	\frac{i}{\sqrt{g_j}} \sum _{q \ne 0}
	\left|
	\frac{\pi}{2qL}
	\right|^{1/2}
	e^{iqx}
	(e^{ |q|v_{j}\tau} \hat{b}^{\dagger}_{j,q}
	-
	e^{-|q|v_{j} \tau} \hat{b}_{j,-q}) \; .
\label{fields}
\end{eqnarray}

The tunneling occurs in the  junctions at the points $x=0$ and $x=d$.
We obtain an effective action, as in the metallic case,
by introducing the auxiliary
field  related to the voltage drop across the junction $\phi
(\tau )$ and performing a gauge transformation which transfers the phase
dependence into the tunneling matrix elements.
After a cumulant expansion  we express the partition function
as a path integral involving  the Euclidean action
\begin{equation}
	S[\phi] = \int_{0}^{\beta} \!\! d\tau\,
	\frac{C}{2e^2} \left(\frac{
	\partial\phi}{\partial\tau} \right)^2
	-\int_{0}^{\beta} \! d\tau\, d\tau'\,
	\, 2
	\alpha(\tau-\tau') \cos[\phi(\tau)-\phi(\tau')]  .
\nonumber
\end{equation}
It has the same form as for noninteracting electrons.
However, the kernel
\begin{equation}
	\alpha (\tau) =  \frac{k_F^2}{2 \pi}
	\mid T\mid ^2 N(0) \left(\frac{\pi /v_ \rho\beta k_F}{\mid
	\sin(\pi\tau/\beta)\mid}\right)^{\eta}
\end{equation}
carries an exponent $\eta$, which is given by the parameters
of the Luttinger liquid
$\eta = 1 + \frac{1}{8} [ g_{\rho} + g_{\sigma} +\frac{4}{g_{\rho}}
 + \frac{4}{g_{\sigma}}]$.
In Fourier space the kernels are proportional to $\omega^{\eta -1}$.
As a result the $I$-$V$ characteristics of a
normal metal-Luttinger liquid junction will also show a power law
dependence proportional to $\mbox{sign}(V) V^{\eta - 1}$.

We point out that the effective action differs from the one
obtained in Ref.~\cite{Kane2} for the model of a Luttinger liquid
with two constriction. The physics associated with Coulomb
blockade is very similar whether we describe the central island as a Luttinger
liquid or as a metal island.
We finally point out that a similar action was discussed in Ref.\cite{ueda}
in connection with shake-up effects close to the Fermi energy due to tunneling,
and in Ref.~\cite{Maura} for junctions coupled to Ohmic baths.

\begin{acknowledgements}
We would like to acknowledge stimulating discussions
with G. Falci, A.D. Zaikin and G. Zimanyi.
This work was supported by the Swiss National Science
Foundation (H.S.) and by the 'Deutsche Forschungsgemeinschaft' as part of
'Sonderforschungsbereich 195'.
\end{acknowledgements}


\begin{thebibliography}{99}
\bibitem{Ave-Lik}
D.V. Averin and K.K. Likharev, in {\it Mesoscopic Phenomena in Solids},
ed. B.L. Altshuler, P.A. Lee and R.A. Webb (Elsevier, Amsterdam,
1991), p. 173

\bibitem{Gra-Dev}
Several review articles are contained in {\it Single Charge
Tunneling}, NATO ASI Series,
Vol. 294,  H. Grabert and M.H. Devoret, eds., New York, Plenum Press (1992)

\bibitem{ZPhB}
For further articles we refer to Z.Phys.B-Condensed Matter {\bf 85} (1991)

\bibitem{Kul-She}
I.O. Kulik and R.I. Schekhter, Sov. Phys. JETP {\bf 41}, 308 (1975)

\bibitem{Schoen-Zai}
G. Sch\"on and A.D. Zaikin, Phys. Rep. {\bf 198}, 237 (1990)

\bibitem{Bee}
C.W.J. Beenakker, Phys. Rev. B {\bf 44}, 1646 (1991)

\bibitem{Ave-Kor-Lik}
D.V. Averin, A.N. Korotkov and K.K. Likharev,
Phys. Rev. B {\bf 44}, 6199 (1991)

\bibitem{Mei-Win-Lee}
Y. Meir, N.S. Wingreen and P.A. Lee, Phys. Rev. Lett. {\bf 66}, 3048 (1991)

\bibitem{Laf-exp}
P. Lafarge, H. Pothier, E.R. Williams, D. Esteve, C. Urbina and
M.H. Devoret, Z. Phys. B - Condensed Matter {\bf 85}, 327 (1991)

\bibitem{Gla-Mat}
L.I. Glazman and K.A. Matveev, Sov. Phys. JETP {\bf 71}, 1031 (1990)

\bibitem{Mat}
K.A. Matveev, Sov. Phys. JETP {\bf 72}, 892 (1991)

\bibitem{Zai}
D.S. Golubev and A.D. Zaikin, Physica B {\bf 203} (1994)

\bibitem{Gra}
H. Grabert, Physica B {\bf 194-196}, 1011 (1994);
Physica B {\bf 203} (1994)

\bibitem{Fal-Scho-Zim}
G. Falci, J. Heinz, G. Sch\"on and G.T. Zimanyi,
Physica B {\bf 203} (1994)

\bibitem{Zwe}
W. Zwerger, preprint

\bibitem{SS}
H. Schoeller and G. Sch\"on, to be publ. in Phys. Rev. B;
Physica B {\bf 203} (1994)

\bibitem{Fey-Ver}
R.P. Feynman and F.L. Vernon, Ann. Phys. (N.Y.) {\bf 24}, 118 (1963)

\bibitem{CLeggett}
A.O. Caldeira and A.J. Leggett, Ann. Phys. (NY) {\bf 149}, 374 (1983)

\bibitem{Eck-Scho-Amb}
U. Eckern, G. Sch\"on and V. Ambegaokar, Phys. Rev. B {\bf 30}, 6419 (1984)

\bibitem{Bic}
N.E. Bickers, Rev. Mod. Phys. {\bf 59}, 845 (1987)

\bibitem{Hew}
A.C. Hewson, {\it The Kondo Problem to Heavy Fermions} (Cambridge Univ.
Press, 1993).

\bibitem{Bar}
S.E. Barnes, J. Phys. F: Metal Phys. {\bf 7}, 2637 (1977);
Phys. Rev. B {\bf 33}, 3209 (1986)

\bibitem{Ram}
J. Rammer, Rev. Mod. Phys. {\bf 63}, 781 (1991)

\bibitem{Los-Schoeller}
D. Loss and H. Schoeller, Physica {\bf 150A} 199 (1988);
J. Stat. Phys. {\bf 54}, 765 (1989); J. Stat. Phys.{\bf 56} 175 (1989)

\bibitem{Haldane}
F.D.M. Haldane, J. Phys. C {\bf 14}, 2585 (1981);
J. S\'olyom, Adv. Phys. {\bf 28}, 201 (1979)

\bibitem{Kane}
C.L. Kane and M.P.A. Fisher, Phys. Rev. Lett. {\bf 68},
1220 (1992); Phys. Rev. B {\bf 46}, 15233 (1992)

\bibitem{Matveev2}
K.A. Matveev and L.I. Glazman, Phys. Rev. Lett. {\bf 70}, 990 (1993)


\bibitem{bruder} For a systematic discussion of higher order terms,
which describe physical processes such as Andreev reflection in
normal-superconductor
junctions or elastic cotunneling, we refer to
C. Bruder, R. Fazio and G. Sch\"on,
Physica B {\bf 203} (1994)

\bibitem{ueda}
M. Ueda and F. Guinea, Z.Phys. {\bf 85}, 413 (1991)

\bibitem{Haldane2}
F.D.M. Haldane, Phys. Rev. Lett. {\bf 47},1840 (1981)

\bibitem{Kane2}
C.L. Kane and M.P.A. Fisher, Phys. Rev. B {\bf 46}, 15233 (1992)

\bibitem{Maura}
M. Sassetti and U. Weiss, this volume
\end{thebibliography}
\end{document}